\documentclass[epj]{svjour}
\usepackage{graphics}
\usepackage{epsfig}
\usepackage{epsf}

\newcommand{\beps}{\mbox{\boldmath$\epsilon$\unboldmath}}

\begin{document}
\title{
Study of temperature dependent atomic correlations in MgB$_{2}$
} 
\subtitle{}
\author{G. Campi\inst{1} \and E.  Cappelluti\inst{2,3}
\and Th. Proffen\inst{4}
\and X. Qiu\inst{5} \and E.~S. Bozin\inst{5} \and S.~J.~L. Billinge\inst{5}
\and S. Agrestini\inst{3} \and N.L. Saini\inst{3} \and A. Bianconi\inst{3}}
\institute{Istituto di Cristallografia, CNR, sezione di
Monterotondo, Area della Ricerca di Roma - Montelibretti, Po. Box
10, 00016 Monterotondo St.(RM), Italy
\and
Istituto dei Sistemi Complessi, CNR-INFM,
v. dei Taurini 19, 00185 Roma, Italy \and
Dipartimento di Fisica, Universit\`{a} di Roma ``La
Sapienza'', P. le Aldo Moro 2, 00185 Roma, Italy
\and
Lujan Neutron Scattering Center, Los Alamos National
Laboratory, Los Alamos, New Mexico 87545, USA
\and
Department of Physics and Astronomy, Michigan State
University, East Lansing, Michigan 48824, USA}
\date{\today/ \mbox{}}
\abstract{
We have studied the evolution with temperature of the local as
well as the average crystal structure of MgB$_2$ using the
real-space atomic pair distribution function (PDF) measured by
high resolution neutron powder diffraction. We have investigated
the correlations of the B-B and B-Mg nearest neighbor pair motion
by comparing, in the wide temperature range from $T=10$~K up to
$T=600$~K, the mean-square displacements (MSD) of single atoms with the
 mean-square relative displacements (MSRD) obtained from the PDF peak
linewidths. The results show that the single atom B and Mg
vibrations are mostly decoupled from each other, with a small
predominance of positive (in phase) correlation factor for both
the B-B and B-Mg pairs. The small positive correlation is almost
temperature independent, in contrast with our theoretical
calculations; this can be a direct consequence of the strong decay
processes of the $E_{2g}$ anharmonic phonons.
}
\maketitle

\section{Introduction}

MgB$_2$ is the simplest system to investigate the quantum mechanism
that allows the formation of a superconducting condensate with
critical temperature $T_c\simeq 40$~K \cite{Nagamatsu} a factor two
higher than in all other known intermetallic superconductors. It was
recently proposed by a few groups
\cite{Agrestini,Imada,Bianconi1,Yamaji,Örd} that the enhancement of
the critical temperature in MgB$_2$ is due to the exchange-like
interband pairing in a multiband superconductor. There is now
experimental evidence \cite{Bouquet,Szabo,Giubileo,Tsuda,Gonnelli}
that MgB$_2$ is the first clear case of a high $T_c$ multiband
superconductor showing two-gaps in the $\sigma$ and $\pi$ bands
respectively, in agreement with the theory.\cite{Liu,Choi}

The characteristic feature of MgB$_2$ is that the electron-phonon
interaction gives a weak pairing in the $\pi$ channel and a strong
pairing in the $\sigma$-channel. Electronic band
calculations,\cite{An,Yildirim,Bohnen,Kong,Boeri} Raman
\cite{Bohnen,Castro} and inelastic neutron scattering experiments
\cite{Shukla} provide evidence of an extremely large deformation
potential for the B bond stretching modes, which gives rise to
strongly anharmonic phonons. This anharmonicity also results in a
structural instability (phonon softening) that affects the dynamics
of the lattice fluctuations and the local structural properties of
the material, as we discuss below. There is now a general agreement
that this strong electron-phonon coupling is mainly driven by the
interaction between electronic carriers in the 2D $\sigma$ band with
boron $p_{x,y}$ character and the zone center $E_{2g}$ phonon
mode.\cite{Liu,Choi,An,Yildirim,Bohnen,Kong} This is reflected in a
Kohn anomaly in the phonon dispersion related to the size of the
small 2D tubular Fermi surfaces.\cite{Shukla} The proximity of the
Fermi level to the Van Hove singularity (VHs) and to the band edge
discloses a new scenario where the large amplitude of the expected
boron zero point lattice fluctuations
\cite{Bouquet,Liu,Yildirim,Boeri,Bianconi2,Boeri2} induces large
fluctuations of the same order of the separation between the VHs,
the gap edge and the Fermi level itself. Although the amplitude of
the lattice fluctuations seems thus to be highly relevant for the
superconductivity in MgB$_2$ there is a lack of
experimental information on this key point. Furthermore, although
the average structure (P6/mmm) of the MgB$_2$ system has been
exhaustively investigated, there is not yet any study of the local
structure, since typical x-ray local probes, such as EXAFS, cannot be
used to study local structure near light atoms.

In view of this, here we employ
high resolution neutron diffraction to obtain the
pair distribution function (PDF) of MgB$_2$.
In this way we investigate
the local as well as the average structure of MgB$_2$,
namely the mean-square displacements (MSD) of single atoms
and the mean-square relative displacements (MSRD).
The comparison of these quantities permits for the first time
to extract the correlation factors
$\rho_{\rm B-B}$, $\rho_{\rm B-Mg}$,
defined below, of the boron-boron and boron-magnesium
pair motions, which are found to
be $\rho_{\rm B-B} \sim 0.1$, $\rho_{\rm B-Mg}\sim 0.1$,
and nearly constant in a wide range of temperature $0$ K $< T < 600$ K.
We also compare the experimental data with
a constant force (CF) model for the phonon dispersion.
We estimate that the phonon frequency renormalization
due to the electron-phonon interaction on the $E_{2g}$ modes
yields a reduction $\Delta \rho_{\rm B-B} \sim -0.03$
in the boron-boron correlation factor.
While the CF model can nicely account for
the zero temperature values
of the single atoms MSD and the correlated MSRD,
the temperature behavior of the correlation factor
is shown to be highly anomalous and
its physical interpretation gives rise to new questions
about our understanding to the local lattice dynamics in this material.

\section{Experimental method and data processing}

Polycrystalline samples of MgB$_{2}$ were synthesized at high
temperature by direct reaction of the elements in a tantalum
crucible under argon atmosphere using pure $^{11}$B isotopes.
Time-of-flight neutron powder diffraction data were collected on the
NPDF diffractometer at the Manuel Lujan, jr., Neutron Scattering
Center (LANSCE) at Los Alamos National Laboratory. The NPDF
diffractometer, with its high neutron flux and backscattering
detector modules has a high resolution,\cite{Proffen2} providing
access to a wide range of momentum transfer with sufficient counting
statistics making it ideal for PDF studies. The powdered samples
were sealed inside extruded cylindrical vanadium containers. These
were mounted on the stage of a cryofurnace with and without
heat-shield for $T<300$~K and $T>300$ K, respectively. The
scattering data from the empty cryofurnace, with and without
heat-shield, an empty container mounted on the cryofurnace, and the
empty instrument were also collected, allowing us to assess and
subtract instrumental backgrounds. The scattering from a vanadium
rod was also measured to allow the data to be normalized for the
incident spectrum and detector efficiencies. We collected each
diffraction spectrum up to the high  momentum transfer of $Q =
40${\AA}$^{-1}$ in 3 hours. The high resolution diffraction spectrum
so obtained presents both the Bragg peaks and the diffuse
scattering. While the Bragg peaks reflect the long-range order of
the crystalline samples, the oscillating diffuse scattering contains
local structural information including the correlated
dynamics.\cite{Jeong,Jeong2} The PDF is obtained from a Fourier
transform of the powder diffraction spectrum (Bragg peaks + diffuse
scattering).\cite{Egami} It consists of a series of peaks, the
positions of which give the distances of atom pairs in the real
space, while the ideal width of these peaks is due both to relative
thermal atomic motion and to static disorder. This permits the study
of the effects of lattice fluctuations on PDF peak widths and yields
information on both the single atom mean-square displacements and
the relative mean-square displacements of atom pairs and their
correlations.

Standard data corrections \cite{Egami}
were carried out using the program PDFGETN.\cite{Peterson1} After
being corrected, the data were normalized by the total scattering
cross section of the sample to yield the total scattering structure
function $S(Q)$. Afterwards, the total scattering structure function
$S(Q)$ is converted to the PDF, $G(r)$, by means of a sine Fourier
transform according to the relation:
\begin{eqnarray}
G(r) &=& 4\pi r (\rho(r)-\rho_0) \nonumber\\ &=&
\frac{2}{\pi}\int_0^\infty Q[S(Q)-1] \sin(Qr)dQ. \label{gr}
\end{eqnarray}
We modeled the PDF using a structural model that takes advantage of
the definition of the radial distribution function RDF $R(r)$,
namely:
\begin{equation}
R(r) =rG(r)+4\pi r^2 \rho_0  = \sum_{i \neq j}\frac{b_i
b_j}{\langle b \rangle^2}\delta (r-r_{ij}),
\end{equation}
where $b_{i}$ is the scattering length of the $i^{th}$ atom,
${\langle b \rangle}$ is the scattering length averaged over the
sample composition,  $r_{ij}=|{\bf r}_i-{\bf r}_j|$ is the distance
between the $i^{th}$ and the $j^{th}$ atoms, and the sums are taken
over all the atoms in the sample. Before being compared to the data,
the calculated $G(r)$ is convoluted with a termination function,
$\sin(Q_{\rm max}r)/r$ to account for the effects of the finite data
collection range. Fundamental lattice information, such as the
average crystal structure, the lattice constants, a scale factor,
and the refined atomic (thermal) displacement parameters, can now be
extracted from the PDF by using the PDF refinement program PDFFIT
\cite{Proffen1} that is based on a least-squares approach to fit the
PDF profile. The average atomic displacement distribution of atom i
along the major axes x, y, and z, $\sigma^2(i_x)$, $\sigma^2(i_y)$,
$\sigma^2(i_z)$, are defined as
\begin{equation}
\sigma^2(i_\alpha) = \langle [{\bf u}_i \cdot \hat{\bf r}_\alpha]^2 \rangle,
\end{equation}
where ${\bf u}_{i}$ is the lattice displacement of atom $i$ from
its average position \cite{Thorpe} and $\hat{\bf r}_\alpha$ is
the unit vector pointing along the direction $\alpha=x,y,z$. Due to
the geometry of these compounds, the two boron atoms for the unit cell
are equivalent and $\sigma^2(i_x)=\sigma^2(i_y)$ for all the atoms,
so that only four parameters were needed namely $\sigma^2({\rm
B}_{xy})$, $\sigma^2({\rm B}_{z})$, $\sigma^2({\rm Mg}_{xy})$ and
$\sigma^2({\rm Mg}_{z})$.

\maketitle
\section{Results and discussion}

\begin{figure}[t]
\epsfxsize 8.0cm \centerline{\epsfbox{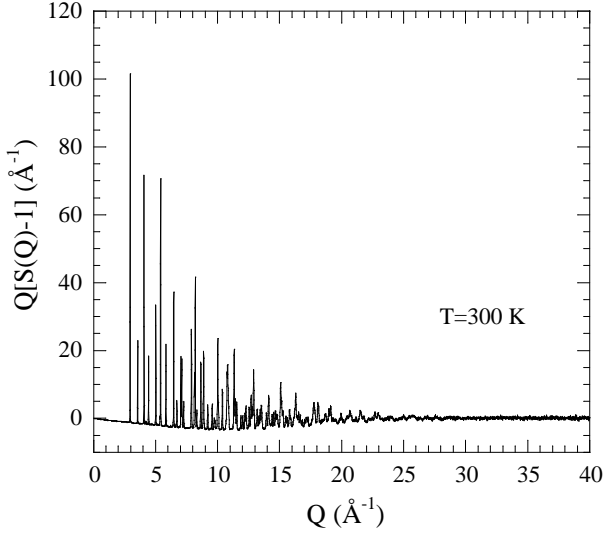}}
\caption{\label{fig:epsart} The reduced structure function
$Q[S(Q)-1]$ for MgB$_2$ measured at $300$~K. }
\end{figure}

\begin{figure}[b]
\epsfxsize 8.0cm \centerline{\epsfbox{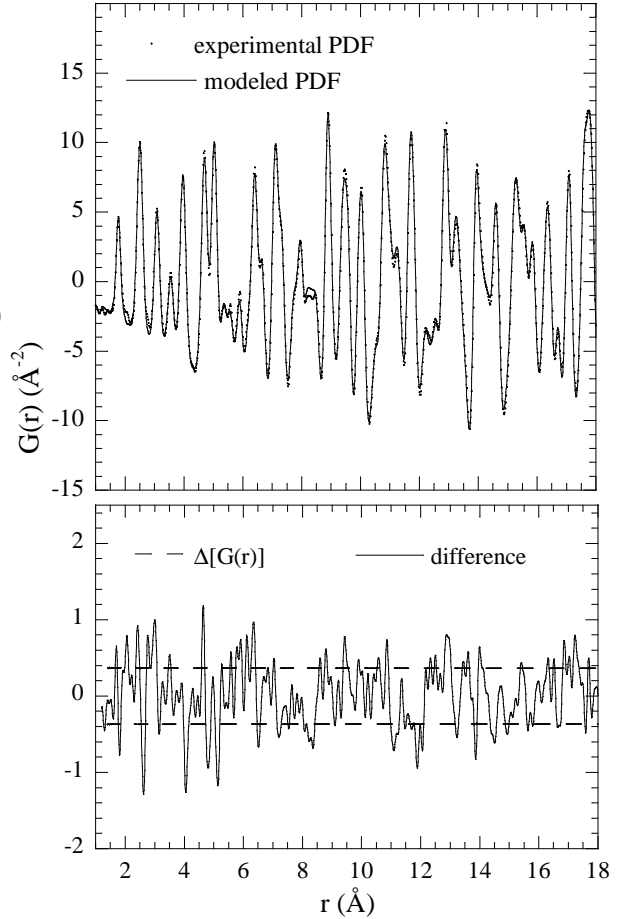}} \caption{\label{sq}
(Upper panel) The PDF $G(r)=4\pi r[\rho(r)-\rho_0]$, obtained from
Eq.~\ref{gr} for MgB$_2$, measured at 300~K (dots), with the
structure refinement curve  obtained by a least-squares approach
(solid line). (Lower panel) The difference curve (solid line) of the
experimental PDF with the modeled fit and the standard deviation on
the data $\pm \Delta$[G(r)] (dashed lines) are shown. We can observe
that most of the fluctuations in the difference curve are within
$\pm \Delta$[G(r)].}
\end{figure}

In Fig.~\ref{fig:epsart}
we show the reduced scattering structure function $Q[S(Q)-1]$
for the MgB$_{2}$ at $T=300$~K, while the corresponding reduced PDF,
$G(r)$, obtained using Eq.~(\ref{gr}), is shown in Fig.~\ref{sq}.
The features of the NPDF diffractometer allowed us to obtain high
quality PDFs as can be noted by inspecting  the modeled fit (solid
line) of the $G(r)$ in the upper panel Fig.~\ref{sq}. This can be
seen also in the lower panel of Fig.~\ref{sq} where most of the
fluctuations in the difference curve (solid line) are within the
standard deviation on the data $\pm \Delta$[G(r)](dashed lines above
and below the difference curve).

The PDF's have been fit over the range $1-18$~{\AA} using
a hexagonal crystal structure (space group P6/mmm), and excellent
fits were obtained at all temperatures. The Bragg peaks are clearly
persistent up to 25~{\AA}$^{-1}$, reflecting both the long-range
order of the crystalline samples and the small amount of positional
(vibrational or static) disorder of the atoms around their average
positions.

\begin{figure}[t]
\epsfxsize 8.4cm \centerline{\epsfbox{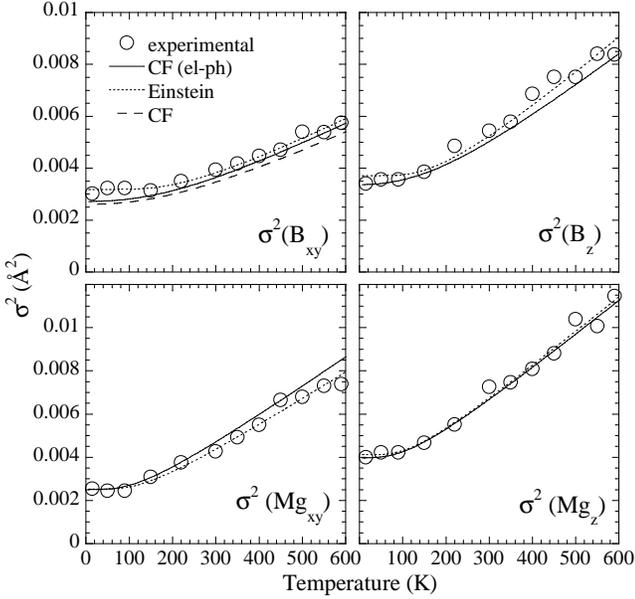}}
\caption{\label{f-gr} Anisotropic in plane $\sigma^2$(B$_{xy}$),
$\sigma^2$(Mg$_{xy}$) and along the c-axis $\sigma^2$(B$_z$),
$\sigma^2$(Mg$_z$) mean-square displacements for the B and Mg atoms,
as a function of the temperature. The open circles represent the
PDFFIT refined values, while the solid and the dotted lines
represent the modeled data in the CF and Einstein model,
respectively. The errorbars in the $\sigma^2$(i$_{\alpha}$) are
smaller than the size of the used symbols (open circles). The dashed
line shows the in-plane boron lattice displacements in the CF model
in the absence of the electron-phonon interaction.}
\end{figure}

The lattice parameters at $T=300$~K for
MgB$_{2}$ ($T_{c} \sim 39$~K) were found to be $a=3.08505(4)$~{\AA},
$c= 3.5218(1)$~{\AA}. The lattice displacements $\sigma^2({\rm
B}_{xy})$, $\sigma^2({\rm B}_{z})$, $\sigma^2({\rm Mg}_{xy})$,
$\sigma^2({\rm Mg}_{z})$  are shown in Fig.~\ref{f-gr} (open
circles).

\subsection{Atomic mean-square displacements}

In order to extract information about the phonon spectrum,
we fit each lattice displacement with a simple Einstein model,
\begin{equation}
\sigma^2(i_\alpha)=\frac{\hbar}{M_i \omega_{i_\alpha}}
\left[
\frac{1}{2}+n[\omega(i_\alpha)]\right]
+\sigma_0^2(i_\alpha),
\label{einstein}
\end{equation}
where $\sigma_0^2(i_\alpha)$ takes into account the static disorder,
$M_i$ is atomic mass of the atom $i$ and $\omega(i_\alpha)$
provided an estimate of the vibrational frequency of the atom $i$
along the direction $\alpha$.
The quantity $n(x)$ is the Bose thermal factor $n[x]=[\exp(x/k_{\rm
B} T)-1]^{-1}$. The values of the fitting parameters,
$\omega_{i_\alpha}$ and $\sigma_0^2(i_\alpha)$, are reported in
Tab.~\ref{t-s2}, and the $\sigma^2$ vs. $T$ fitting curves are
represented by dotted lines in Fig.~\ref{f-gr}.
The different values of $\omega_{i_\alpha}$ represent the different
energy range of the phonon spectra associated with the boron and magnesium
in-plane and out-of-plane lattice vibrations, and they are in good
agreement with the corresponding spectra reported in Ref. \cite{osborn}.

\begin{table}
\begin{tabular}{|c|c|}
\hline\hline
$\omega({\rm B}_{xy})$  (K) & $703 \pm 10$  \\
\hline
$\omega({\rm B}_{z})$  (K) & $572 \pm 15$  \\
\hline
$\omega({\rm Mg}_{xy})$  (K) & $398 \pm 6$  \\
\hline
$\omega({\rm Mg}_{z})$  (K) & $350 \pm 9$  \\
\hline & \\
\hline
$\sigma_0^2({\rm B}_{xy})$ (\AA$^2$)
& $\le 5\cdot 10^{-5}$  \\
\hline $\sigma_0^2({\rm B}_{z})$ (\AA$^2$)
& $\le 9\cdot 10^{-5}$  \\
\hline $\sigma_0^2({\rm Mg}_{xy})$ (\AA$^2$)
& $\le 7\cdot 10^{-5}$  \\
\hline $\sigma_0^2({\rm Mg}_{z})$ (\AA$^2$)
& $0.0013 \pm 0.0003$  \\
\hline\hline
\end{tabular}
\caption{``Effective'' phonon frequencies $\omega_{i_\alpha}$
and static disorder contributions $\sigma_0^2(i_\alpha)$
for the in-plane and out-of-plane B and Mg displacements
as obtained by the Einstein model fit [Eq. (\ref{einstein})]
of $\sigma^2(i_\alpha)$.}
\label{t-s2}
\end{table}

Quite surprising, we find that, apart from some amount of disorder
in $\sigma_0^2({\rm Mg}_{z})$, this simple four-peak Einstein model
seems to describe quite well the lattice vibrations in this
compound. This is quite intriguing because phonon frequency
dispersions are expected to be notably different from dispersionless
Einstein models in these materials. Moreover, the Einstein fits were
obtained by simply using an effective atomic mass $M_{\rm Mg} =
24.3$ a.m.u. for $\sigma^2({\rm Mg})$ and $M_{\rm B} = 10.81$ a.m.u.
for $\sigma^2({\rm B})$. This means that we are implicitly assuming
that magnesium and boron vibrations are totally decoupled from each
other. These two-parameter fits reproduce the data very well for
each of the motions considered.
Non-independent boron and magnesium vibrations would lead to
deviations from Einstein behavior that are not evident in the data
suggesting that this approximation is reasonable in this system.  We
return to this point below.

To gain further insight on this issue we introduce a constant force
(CF) shell model for the dynamical matrix.
We neglect for the moment the effects of the electron-phonon
interaction, which leads to renormalized phonon frequencies
and anharmonicity, and we assume
the lattice dynamics to be harmonic (we shall discuss later and more
specifically the role of the electron-phonon interaction
and possible effects of the
anharmonicity of the $E_{2g}$ phonon mode). This will enable us
to evaluate eigenvectors $\hat{\beps}_{{\bf q},\mu}$ and eigenvalues
$\omega_{{\bf q},\mu}$ of the lattice modes for each point of the
phonon Brillouin zone. The phonon contribution to the lattice
displacements $\sigma^2(i_\alpha)$
[$\sigma^2(i_\alpha)=\sigma^2_{\rm
ph}(i_\alpha)+\sigma^2_0(i_\alpha)$] will be thus obtained
as:\cite{Chung}
\begin{equation}
\sigma^2_{\rm ph}(i_\alpha) =
\frac{\hbar}{N} \sum_{{\bf q},\mu}
\frac{|\epsilon_{{\bf q},\mu}^{i\alpha} |^2}
{M_i \omega_{{\bf q},\mu}}
\left[
\frac{1}{2}+n(\omega_{{\bf q},\mu})
\right],
\label{e-s2}
\end{equation}
where $\epsilon_{{\bf q},\mu}^{i\alpha}$ is the component of the
eigenvector $\hat{\beps}_{{\bf q},\mu}$ concerning to the
displacement of the $i$ atom along the $\alpha$ direction and $N$ is
the total number of ${\bf q}$-points considered in the phonon
Brillouin zone. From a general point of view, since $\sigma^2_{\rm
ph}(i_\alpha)$ involves an integral over the whole Brillouin zone
and over all the phonon branches, it will not be sensitive to the
fine details of the phonon dispersion but only to its gross
features. For this reason, and in order to preserve the simplicity
of our analysis, we limit ourselves to consider only four elastic
springs, $\phi$, $\chi$, $\kappa$ and $\psi$, connecting,
respectively, in-plane B-B nearest neighbors, out of plane B-B
nearest neighbors, out-of-plane B-Mg nearest neighbors, and in-plane
Mg-Mg nearest neighbors. Each elastic spring is specified by its
tensor components (ex.: $\phi_r$, $\phi_\|$, $\phi_\perp$),
corresponding respectively to the lattice displacements along the
radial (bond-stretching) direction and along the
in-plane and out-of-plane tangential (bond-bending) directions.
In MgB$_2$ we choose
the constants $\phi_r$, $\phi_\|$, $\phi_\perp$, $\chi_r$,
$\chi_\|$, $\chi_\perp$, $\kappa_r$, $\kappa_\|$, $\kappa_\perp$,
$\psi_r$, $\psi_\|$, $\psi_\perp$, to fit the local-density
functional (LDA) phonon dispersion of Ref. \cite{Bohnen} along
the high-symmetry points of the Brillouin zone. Since the constant
force model is meant to reproduce the bare phonon dispersion, we
deliberately did not include in the fit procedure the $E_{2g}$
phonon frequencies at the zone center $\Gamma$, $A$, which are known
to be strongly affected by the el-ph interaction. The elastic
constants obtained from this fitting procedure are reported in
Table~\ref{t-par}.
\begin{table}
\begin{tabular}{|c|c|c|}
\hline
$\phi_r$ (eV/{\AA}$^2$) & $\phi_\|$ (eV/{\AA}$^2$) &
$\phi_\perp$ (eV/{\AA}$^2$)  \\
\hline
$12,45$  & $49.80$  & $21.17$  \\
\hline\hline
$\chi_r$  (eV/{\AA}$^2$)& $\chi_\|$  (eV/{\AA}$^2$) &
 $\chi_\perp$  (eV/{\AA}$^2$)\\
\hline
$0.0$ & $16.6$ & $0.0$ \\
\hline\hline
$\kappa_r$ (eV/{\AA}$^2$) & $\kappa_\|$  (eV/{\AA}$^2$) &
$\kappa_\perp$  (eV/{\AA}$^2$)\\
\hline
$0.84$ & $1.50$ & $9.14$ \\
\hline\hline
$\psi_r$  (eV/{\AA}$^2$)& $\psi_\|$ (eV/{\AA}$^2$) &
$\psi_\perp$ (eV/{\AA}$^2$) \\
\hline
$0.0$ & $9.34$ & $0.42$ \\
\hline
\end{tabular}
\caption{Force constant parameters reproducing the bare phonon
dispersion in MgB$_2$ in the absence of electron-phonon
interaction.} \label{t-par}
\end{table}

As mentioned in the introduction,
the $E_{2g}$ phonon modes close to the $\Gamma$
and $A$ points are expected to be strongly affected by the
interaction with the almost 2D parabolic $\sigma$ bands, giving rise
to a remarkable softening of the $E_{2g}$ phonon frequencies for
$|{\bf q}|\le 2k_{\rm F}$, where $k_{\rm F}$ is the Fermi vector of
the $\sigma$ bands.\cite{An,Yildirim,Bohnen,Kong,Shukla} We include
these effects through the self-energy renormalization of the phonon
frequencies $\Omega_{E_2g}^2({\bf q})= \omega_{E_2g}^2({\bf q}) -(4
N_\sigma g^2 f_{\rm anharm}/M_{\rm B}) \Pi_{\rm 2D}({\bf q})$, where
$N_\sigma$ is density of states of the $\sigma$ bands per spin and
per band, $g$ the electron-phonon matrix element between
$\sigma$-band electrons and the $E_{2g}$ phonon mode at the zone
center and $f_{\rm anharm}$ is a dimensionless factor accounting for
the anharmonic hardening of the $E_{2g}$ phonon modes due to the
electron-phonon coupling itself. Moreover the factor $4$ takes into
account the spin and band degeneracy and $\Pi_{\rm 2D}({\bf q})$ is
the two-dimensional Lindhardt function $\Pi_{\rm 2D}(x)=
\theta(1-x)+ \theta(x-1) [x-\sqrt{x^2-1}]/x^4$, with $x=|{\bf
q}|/2k_{\rm F}$. We take, from first-principle
calculations,\cite{An,Yildirim,Boeri} $f_{\rm anharm} = 1.25$,
$N_\sigma = 0.075$ states/($\mbox{eV} \cdot \mbox{spin} \cdot
\mbox{cell}$) $g = 12$ eV/\AA\, and $k_{\rm F} \simeq \pi/12 d_{\rm
B-B}$, where $d_{\rm B-B}$ is the boron-boron distance.

\begin{figure}[t]
\centerline{\epsfig{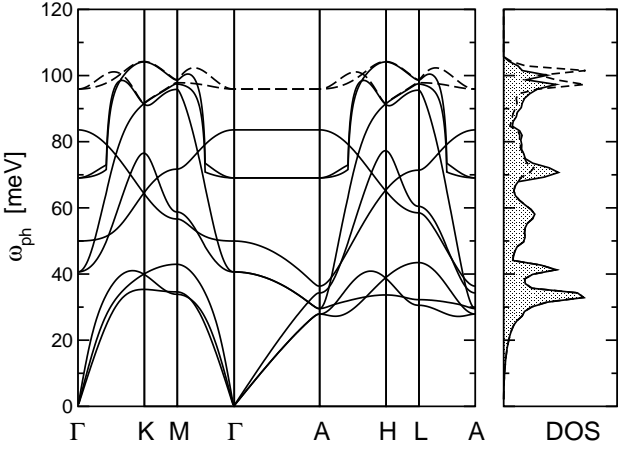}}
\caption{\label{f-disp} Phonon dispersion and phonon density of
states for MgB$_2$ evaluated within the CF model. Dashed lines
represent the same quantities without taking into account the
electron-phonon interaction.}
\end{figure}

The phonon
dispersion and phonon density of states (PDOS) of our constant force
model are shown in Fig. \ref{f-disp},
in good agreement with the LDA calculations of Ref.
\cite{Bohnen}. For comparison, the dashed lines represent the bare
phonon dispersion in the absence of el-ph interaction and the
corresponding PDOS.
Most striking is a partial shift in spectral
weight from 100~meV to 70~meV when the el-ph coupling is turned on,
due to the softening of the flat $\Gamma - A$ band originating from
the $E_{2g}$ phonons.

The lattice displacements for each atom and for each direction can
now be evaluated directly by means of Eq.~(\ref{e-s2}). The results
are also shown in Fig.~\ref{f-gr} along with the data extracted by
the PDF analysis and with the simple Einstein fits. In spite of the
crude simplification of the phonon dispersion, the agreement with
the PDF data is remarkable. Just as in the Einstein fits, we find
that a constant shift $\sigma^2_0({\rm Mg}_{z}) \simeq 0.0013$
{\AA}$^2$ due to intrinsic disorder is needed to account for the
magnesium out-of-plane lattice vibration while the contribution of
the disorder for the other modes is found to be negligible. Note
that the agreement between the data and solid lines was obtained
with {\it no adjustable parameters}. The values used in the constant
force models in Table~\ref{t-par} were indeed fit to reproduce the
theoretical LDA calculation and not the PDF data and the only other
parameter, the static disorder parameter in $\sigma^2_0({\rm
Mg}_{z})$, was the same value as was used in the Einstein fits. The
agreement with the data and the Einstein model fits is also very
good suggesting that boron and magnesium lattice vibrations, as well
as in-plane and out-of-plane lattice vibrations, are on average
independent each other.

To point out the explicit role of the electron-phonon interaction,
we show also in the left-upper panel of Fig.~\ref{f-gr} the
$\sigma^2_0({\rm B}_{xy})$ lattice displacements in the absence of
the el-ph frequency renormalization. Since only $E_{2g}$ phonons are
coupled, only the $\sigma^2_0({\rm B}_{xy})$ lattice displacements
result modified. We note that the inclusion of electron-phonon
interaction effects leads to a slight increase of the amount of the
boron in-plane lattice displacements, with a better agreement with
the experimental data. The increase of $\sigma^2_0({\rm B}_{xy})$ is
easily understandable as due to the softening of the $E_{2g}$ phonon
mode. On the other hand, since the electron-phonon renormalization
effects are restricted to a small region $\sqrt{q_x^2+q_y^2}\le
2k_{\rm F}$ of the whole Brillouin zone, the impact of the el-ph
coupling on the total amount of the lattice displacements
$\sigma^2_0({\rm B}_{xy})$ is relatively weak. As we are going to
see, the effects of the electron-phonon interaction are more
apparent in the correlated pair motion.

\subsection{Correlations in the B-B and B-Mg atomic pairs motion}

Above we showed that the average uncorrelated thermal motions
(equivalent to the Debye-Waller factor
in crystallography) measured from MgB$_2$ are well explained by
harmonic models with independent boron and magnesium motions.
A strength of the PDF technique is that it is sensitive to correlations
in the atomic dynamics that contain some additional details
about the underlying
interatomic potentials.\cite{Thorpe,Jeong,Graf,Jeong2}  Here we
explore the motional correlations in the MgB$_2$ PDF data.

The Gaussian width $\sigma_{ij}$ of the PDF peaks
is directly related to the mean-square relative displacement of
atomic pairs projected onto the vector joining the atom
pairs.\cite{Thorpe} Explicitly,
\begin{equation}
\sigma^2_{ij} = \langle [({\bf u}_i-{\bf u}_j)\cdot
\hat{\bf r}_{ij}]^2 \rangle ,
\label{e-corr}
\end{equation}
where ${\bf u}_i$ and ${\bf u}_j$ are the lattice displacements of
atoms $i$ and $j$ from their average positions, $\hat{\bf r}_{ij}$
is the unit vector connecting atoms $i$ and $j$, and where the
angular brackets indicate an ensemble average.\cite{Thorpe}

\begin{figure}
\input{epsf}
\epsfxsize 8.4cm \centerline{\epsfbox{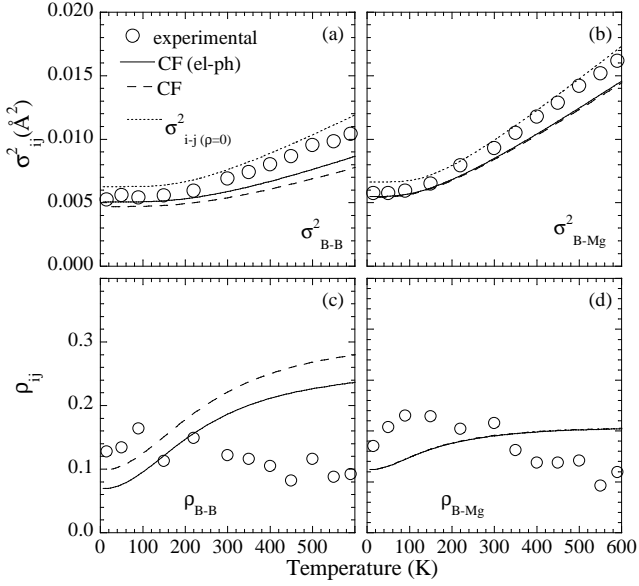}}
\caption{\label{f-sigma2} Mean-square relative lattice displacements
$\sigma^2_{\rm B-B}$ (a), $\sigma^2_{\rm B-Mg}$ (b), as extracted
from the width of the PDF peaks and the corresponding correlation
function $\rho_{\rm B-B}$ (c), $\rho_{\rm B-Mg}$ (d) (open circles).
Solid lines represent the same quantities evaluated within the CF
model, while dotted lines in the upper panels show the predicted
relative lattice displacements $\sigma^2_{\rm B-B}$, $\sigma^2_{\rm
B-Mg}$ for completely uncorrelated lattice motion $\rho_{\rm
B-B}=0$, $\rho_{\rm B-Mg}=0$. The dashed lines in the upper panels
represent the relative lattice displacement $\sigma^2_{\rm B-B}$ and
$\sigma^2_{\rm B-Mg}$ evaluated within the CF model in the absence
of the electron-phonon interaction. }
\end{figure}

In our analysis, we focus on the width of the nearest neighbor
boron-boron PDF peak,
($\sigma^2_{\rm B-B}$), and on the nearest neighbor magnesium-boron
peak, ($\sigma^2_{\rm B-Mg}$).  These are well resolved
single-component peaks in the
PDF whose width directly yields correlated
dynamical information.~\cite{Jeong2}  The Gaussian widths
$\sigma^2_{\rm B-B}$, $\sigma^2_{\rm B-Mg}$, as measured by the PDF
data are shown in the upper panels of Fig.~\ref{f-sigma2} (open
circles)
along with the same quantity
evaluated within the CF model (solid line) according to the relation:
\begin{eqnarray}
\sigma^2_{ij} &=& \frac{\hbar}{N} \sum_{\bf q,\mu}
\left[\frac{1}{2}+n(\omega_{\bf q,\mu})\right]
\Bigg\{
\frac{|\hat{\beps}^i_{{\bf q},\mu} \cdot \hat{\bf r}_{ij}|^2}
{M_i\omega_{\bf q,\mu}}
+\frac{|\hat{\beps}^j_{{\bf q},\mu} \cdot \hat{\bf r}_{ij}|^2}
{M_j\omega_{\bf q,\mu}}
\nonumber\\
&&
-\frac{2\mbox{Re}\left[
(\hat{\beps}^i_{{\bf q},\mu} \cdot \hat{\bf r}_{ij})
(\hat{\beps}^{j*}_{{\bf q},\mu} \cdot \hat{\bf r}_{ij})
\mbox{e}^{i{\bf q}\cdot {\bf r}_{ij}}
\right]}{\omega_{\bf q,\mu}\sqrt{M_iM_j}}
\Bigg\}.
\end{eqnarray}
In this latter case in the right panel for $\sigma^2_{\rm B-Mg}$ we have
also added a small contribution of the local lattice displacements
due to the disorder (see below for more details).

Here we first note that, while the zero temperature values of
$\sigma^2_{\rm B-B}$, $\sigma^2_{\rm B-Mg}$  are well reproduced by
the CF model, this model is less good than was the case for the
uncorrelated motions, especially at high temperatures.

The temperature dependence of the single-atom motions was well
explained by the model, but not those of the correlated
$\sigma^2_{\rm ij}$, which suggests that the model is not capturing
some aspect of the motional correlations. Introducing the
electron-phonon coupling into the model improves the agreement
slightly, with a larger effect observed on $\sigma^2_{\rm B-B}$, but
this does not explain all of the discrepancy. In each case, the pair
correlation peaks broaden more quickly in the data than in the
model. We note that the models are harmonic. Even in the case where
we have introduced electron-phonon coupling, harmonic spring
constants have been obtained from a fit to the LDA bands and we
speculate that the discrepancy at high temperature is is a result of
anharmonicity in the boron motion.

As a general consideration, we would like to stress once more that
while $\sigma^2(i_\alpha)$ probes the absolute magnitude
of the single atom mean-square displacement, $\sigma^2_{ij}$ provides
information about the {\em correlation} between the lattice
displacements of atom pairs.
Let us consider for instance the case
of $\sigma^2_{\rm B-B}$ which involves only boron in-plane lattice
fluctuations. We can identify three limiting behaviors for
this quantity: $i$) perfectly in-phase lattice motion: $ii$)
perfectly opposite-phase motion; $iii$) completely independent
motion. In the first case it is easy to see that $\sigma^2_{\rm
B-B}=0$, while $\sigma^2_{\rm B-B}=2 \sigma^2({\rm B}_{xy})$ when
the nearest neighbor boron lattice displacements are uncorrelated,
and $\sigma^2_{\rm B-B}=4 \sigma^2({\rm B}_{xy})$ when they have
opposite phase.

To formalize this we
rearrange Eq. (\ref{e-corr}) as
\begin{equation}
\sigma^2_{ij} = \langle [({\bf u}_i\cdot\hat{\bf r}_{ij}]^2
\rangle + \langle [({\bf u}_j\cdot\hat{\bf r}_{ij}]^2
\rangle -2\langle ({\bf u}_i\cdot\hat{\bf r}_{ij}) \cdot
({\bf u}_j\cdot\hat{\bf r}_{ij}) \rangle.
\end{equation}
Here the first two terms are related to mean-square thermal
displacement of atoms $i$ and $j$ projected along $\hat{\bf
r}_{ij}$, while the third term is a displacement correlation
function, which carries information about the motional correlations.
It is now useful to quantify the degree of correlation by
introducing the dimensionless correlation parameter $\rho_{ij}$
defined as:\cite{Chung,Jeong,Jeong2,Booth}
\begin{equation}
\sigma^2_{ij} = \sigma^2(i_j) + \sigma^2(j_i)
-2\sigma(i_j)\sigma(j_i)\rho_{ij},
\label{e-rho}
\end{equation}
where $\sigma^2(i_j)= \langle [({\bf u}_i\cdot\hat{{\bf r}}_{ij}]^2
\rangle$. Positive values of $\rho > 0$ describe a situation where
the atoms move in phase, so that the resulting value of
$\sigma^2_{ij}$ is smaller than for the uncorrelated case. On the
other hand, a predominance of opposite phase atomic vibrations
should result in $\rho < 0$ and in a PDF peak width $\sigma^2_{ij}$
larger than the uncorrelated case.

It is important to note that the correlation function
$\rho_{ij}$ in Eq.~(\ref{e-rho}) expresses the degree of correlation
between the {\em total} atomic displacements. In the presence
of two different sources of lattice displacements
(phonons and disorder), it is more convenient to split
$\sigma^2_{ij}$ into a phonon and a disorder contribution.
Assuming the local lattice displacements due to the disorder
to be uncorrelated, we can write thus:
\begin{eqnarray}
\sigma^2_{ij} &=& \sigma^2_{\rm ph}(i_j) + \sigma^2_{\rm ph}(j_i)
-2\sigma_{\rm ph}(i_j)\sigma_{\rm ph}(j_i)\rho_{ij}
\nonumber\\
&&
+\sigma^2_0(i_j) + \sigma^2_0(j_i),
\label{e-rho12}
\end{eqnarray}
where $\rho_{ij}$ represents now only the correlation
between {\em phononic} lattice displacements.

Using Eq.~(\ref{e-rho12})
the correlation parameter
can be calculated from the total width of the
PDF peak as
\begin{equation}
\rho_{ij} = \frac{\sigma^2_{\rm ph}(i_j) + \sigma^2_{\rm ph}(j_i)
+\sigma^2_0(i_j) + \sigma^2_0(j_i)
-\sigma^2_{ij}}{2\sigma_{\rm ph}(i_j)\sigma_{\rm ph}(j_i)}.
\label{e-rho2}
\end{equation}
Finally, the projected atomic mean-square displacements
$\sigma^2(i_j)$ (lattice vibrations along the pair
$\hat{{\bf r}}_{ij}$ direction) can be related to $\sigma^2_{i_\alpha}$
(lattice vibrations along the Cartesian axes) by simple geometrical
considerations.  We have thus
$\sigma^2_{\rm ph}({\rm B}_{\rm B})=\sigma^2_{\rm ph}({\rm B}_{xy})$,
$\sigma^2_{\rm ph}({\rm B}_{\rm Mg})=[4R^2\sigma^2_{\rm ph}({\rm B}_{xy})
+3\sigma^2_{\rm ph}({\rm B}_{z})]/(4R^2+3)$,
$\sigma^2_{\rm ph}({\rm Mg}_{\rm B})=[4R^2\sigma^2_{\rm ph}({\rm Mg}_{xy})
+3\sigma^2_{\rm ph}({\rm Mg}_{z})]/(4R^2+3)$, where $R=a/c=0.88$ and where
$a$ and $c$ are the in-plane and out-of-plane lattice constants.
Similar relations hold true for the disorder contributions.

The phonon correlation factor $\rho_{ij}$ as extracted from the PDF
data $\sigma^2_{\rm B-B}$, $\sigma^2_{\rm B-Mg}$, and from the
single atom mean-square lattice displacements $\sigma^2(i_\alpha)$
is shown in the lower panels of Fig.~\ref{f-sigma2} (open circles),
together with the corresponding correlation factor predicted by the
CF model (solid lines). In order to extract the experimental value
of $\rho_{\rm B-Mg}$ we have taken into account a slight magnesium
disorder along the $c$ axis, $\sigma^2_0({\rm Mg}_{z}) =
0.0013$~{\AA}$^2$, in agreement with the previous analysis of the
mean square absolute displacements $\sigma^2(i_\alpha)$. We find a
positive correlation factor for both $\rho_{\rm B-B} \sim 0.1$ and
$\rho_{\rm B-Mg} \sim 0.1$, indicating a slight predominance of the
in-phase B-B and B-Mg lattice displacements in this experimental
probe. Positive values of $\rho_{ij}$ are commonly reported in a
variety of materials. The intuitive explanation is that the in-phase
phonon modes (acoustic, low optical branch modes) are generally less
stiff than the opposite-phase optical ones. Our reported values of
$\rho_{\rm B-B}, \rho_{\rm B-Mg}\sim 0.1$ are however much smaller
than the correlation factors commonly found in other covalently
bonded materials.\cite{Chung,Jeong,Jeong2,Booth} This observation
points out once more that all the motions are decoupled and the
atoms are behaving largely like independent oscillators.

Another anomalous feature of MgB$_2$ pointed out by this analysis is
the lack of a temperature dependence for the correlation factors
$\rho_{ij}$ as compared with the CF model and with the standard
behavior of other common materials.\cite{Jeong2} An increase of
$\rho_{ij}$ as function of temperature is indeed observed in many
covalently bonded systems and it is essentially due to the fact that
the thermal population of the low frequency in-phase phonon modes is
larger than the high frequency out-of-phase phonon modes. The lack
of this temperature dependence in our measurements can maybe be
attributed to the anharmonic character of the high frequency
($E_{2g}$) B-B modes and it represents an interesting anomaly in
this material whose physical interpretation can shed interesting
light on the lattice dynamics in MgB$_2$. Further work on this
subject is required. As a final point, we can quantify in our model
the role of the electron-phonon coupling on the correlation factors.
As shown in Fig. \ref{f-sigma2} the inclusion of the electron-phonon
interaction, which leads to a partial softening of the out-of-phase
$E_{2g}$ in-plane boron displacements, is reflected in a significant
reduction $\Delta \rho_{\rm B-B} \sim - 0.03$ of the correlation
factor $\rho_{\rm B-B}$ while a negligible effect is found on
$\rho_{\rm B-Mg}$.

\section{Conclusions}

In this work we have investigated the local lattice properties of
MgB$_2$ paying special attention on the lattice dynamics and the
correlations in the B-B and B-Mg first neighbor atomic pair motion.
We have used the real space PDF obtained from high resolution
neutron diffraction to study the effects of the lattice vibrations
on the PDF peak widths. The PDF peaks in well ordered crystals such
as the present case yield important information about the underlying
atomic potentials through the correlated local lattice dynamics. The
data have been modeled using both a multi-parameter constant force
model and a simple Einstein model. We have found that the constant
force model as well as the Einstein one reproduce the average
features of the lattice vibrations. This agreement suggests that
boron and magnesium displacements, both in-plane and out-of-plane,
are mostly independent of each other. The analysis of the PDF peak
linewidths permits to evaluate the correlation for both the nearest
neighbor B-B and B-Mg atomic pairs. We find a small positive
correlation factor $\rho_{\rm B-B} \sim 0.1$ and $\rho_{\rm B-Mg}
\sim 0.1$, nearly temperature independent, indicating a weakly
prevalent in-phase relative atomic motion. These results are in
contrast with CF model which predicts correlation factors increasing
with the temperature. This discrepancy supports the idea that
anharmonic effects and strong decay processes for the $E_{2g}$ B
bond stretching modes are present, presumably due to the strong
electron-phonon coupling.

\section*{Acknowledgements}
This work is supported by MIUR in the frame of the
project Cofin 2003 ``Leghe e composti intermetallici: stabilit\`{a}
termodinamica, propriet\`{a} fisiche e reattivit\`{a}" on the
``synthesis and properties of new borides" and by European project
517039 ``Controlling Mesoscopic Phase Separation" (COMEPHS) (2005).
E.C. acknowledges in addition funding from the FIRB project
RBAUO17S8R of MIUR.  Work in the Billinge-group is supported by
NSF through grant DMR-0304391.  The NPDF diffractometer at the Lujan Center,
Los Alamos National Laboratory,
was funded by DOE through contract W-7405-ENG-36.


\begin{thebibliography}{99}

\bibitem{Nagamatsu}
J. Nagamatsu, N. Nakagawa, T. Muranaka, Y. Zenitali, and J.
Akimitsu, Nature \textbf{410}, 63 (2001).
\bibitem{Agrestini}
S. Agrestini, D. Di Castro, M. Sansone, N. L. Saini, A. Saccone, S.
De Negri, M. Giovannini, M. Colapietro, and A. Bianconi, J. of Phys.:
Cond. Matter {\bf 13}, 11689 (2001).
\bibitem{Bianconi1}
A. Bianconi, D. Di Castro, S. Agrestini, G. Campi, N.L. Saini, A.
Saccone, S. De Negri, and  M. Giovannini, J. Phys.: Condens. Matter {\bf
13}, 7383 (2001).
\bibitem{Imada}
M. Imada, J. Phys. Soc. Jpn. {\bf 70}, 1218 (2001).
\bibitem{Yamaji}
K. Yamaji, J. Phys. Soc. Jpn. {\bf 70}, 1476 (2001).
\bibitem{Örd}
T. \"Ord and N. Kristoffel, Physica C {\bf 370}, 17 (2002).
\bibitem{Bouquet}
F. Bouquet,  R.A. Fisher, N.E. Phillips, D.G. Hinks, and  J.D.
Jorgensen, Phys. Rev. Lett. {\bf 87}, 047001 (2001).
\bibitem{Szabo}
P. Szabo, P. Samuely, J. Kacmarcik, Th. Klein,  J. Marcus, D.
Fruchart, S. Miraglia, C.Marcenat, and A.G.M.Jansen,
Phys. Rev. Lett {\bf 87}, 137005 (2001).
\bibitem{Giubileo}
F. Giubileo, D. Roditchev, M. Sacks, R. Lassey, D.X. Thanh, and J.
Klein, Phys. Rev. Lett. \textbf{87}, 177008 (2001).
\bibitem{Tsuda}
S. Tsuda, T. Yokoya, Y. Takano, H. Kito, A. Matsushita, F. Yin, J.
Itoh, H. Harima, and S. Shin, Phys. Rev. Lett. {\bf 91}, 127001
(2003).

\bibitem{Gonnelli}
R.S. Gonnelli, D. Daghero, G.A. Ummarino, V.A. Stepanov,
J. Jun, S.M. Kazakov, and J. Karpinski,
Phys. Rev. Lett. {\bf 89}, 247004 (2003).

\bibitem{Liu}
A.Y. Liu,  I.I. Mazin, and J. Kortus,
Phys. Rev. Lett {\bf 87}, 087005 (2001).

\bibitem{Choi}
H.J. Choi, D. Roundy, H. Sun, M.L. Cohen, and S.G. Louie,
Phys. Rev. B {\bf 66}, 020513 (2002);
{\em ibid.} Nature, {\bf 418}, 758 (2002).

\bibitem{An}
J.M. An and W.E. Pickett, Phys. Rev. Lett. \textbf{86}, 4366 (2001).

\bibitem{Yildirim}
T. Yildirim, O. G\"ulseren, J.W. Lynn, C.M. Brown, T.J. Udovic, Q.
Huang, N. Rogado, K.A. Regan, M.A. Hayward, J.S. Slusky, T. He, M.K.
Haas, P. Khalifah, K. Inumaru, and R.J. Cava, Phys. Rev. Lett. {\bf
87}, 037001 (2001).

\bibitem{Bohnen}
K.-P. Bohnen, R. Heid, and B. Renker, Phys. Rev. Lett. {\bf 86},
5771 (2001).

\bibitem{Kong}
Y. Kong, O.V. Dolgov, O. Jepsen, and O.K. Andersen, Phys. Rev. B
{\bf 64}, 020501 (2001).

\bibitem{Boeri}
L. Boeri, G.B. Bachelet, E. Cappelluti, and L. Pietronero, Phys.
Rev. B {\bf 65}, 214501 (2002).

\bibitem{Castro}
D. Di Castro, S. Agrestini, G. Campi, A. Cassetta, M. Colapietro,
A.Congeduti, A. Continenza, S. De Negri, M. Giovannini, S. Massidda,
M. Nardone, A. Pifferi, P. Postorino, G. Profeta, A. Saccone, N. L.
Saini, G. Satta, and A. Bianconi,
 Europhys.Lett. {\bf 58}, 278 (2002).

\bibitem{Shukla}
A. Shukla, M. Calandra, M. d'Astuto, M. Lazzeri, F. Mauri, C.
Bellin, M. Krisch, J. Karpinski, S.M. Kazakov, J. Jun, D. Daghero,
and K. Parlinski, Phys. Rev. Lett. {\bf 90}, 095506 (2003).

\bibitem{Bianconi2}
A. Bianconi, S. Agrestini, D. Di Castro, G. Campi, G. Zangari, N.L.
Saini, A. Saccone, S. De Negri, M. Giovannini, G. Profeta, A.
Continenza, G. Satta, S. Massidda, A. Cassetta, A. Pifferi and M.
Colapietro, Phys. Rev. B {\bf 65}, 174515 (2002).

\bibitem{Boeri2}
L. Boeri, E. Cappelluti, and L. Pietronero, Phys.
Rev. B {\bf 71}, 012501 (2005).

\bibitem{Egami}
T. Egami and S.J.L. Billinge, {\em Underneath the Bragg peaks:
structural analysis of complex materials} (Pergamon Press, Oxford,
2003).
\bibitem{Proffen2}
Th. Proffen, T. Egami, S.J.L. Billinge, A.K. Cheetham, D. Louca and
J.B. Parise, Appl. Phys. A-Mater. Sci. Process. {\bf 74}, S163 (2002)
\bibitem{Peterson1}
P.F. Peterson, M. Gutmann, Th. Proffen, and S.J.L. Billinge, J.
Appl. Crystallogr. \textbf{33}, 1192 (2000).
\bibitem{Proffen1}
Th. Proffen and S.J.L. Billinge, J. Appl. Crystallogr. \textbf{32},
572 (1999).

\bibitem{Thorpe}
 M. F. Thorpe, V. A. Levashov, M. Lei and S. J. L. Billinge, In
{\it From semiconductors to proteins: beyond the average structure},
Edited by S. J. L. Billinge and M. F. Thorpe, pp. 105,
(Kluwer/Plenum, New York, 2002).

\bibitem{osborn}
R. Osborn, E.A. Goremychkin, A.I. Kolesnikov, and D.G. Hinks,
Phys. Rev. Lett. {\bf 87}, 017005 (2001).


\bibitem{Jeong}
I.-K. Jeong, Th. Proffen, F. Mohiuddin-Jacobs, and S.J.L. Billinge,
J. Phys. Chem. A {\bf 103}, 921 (1999).
\bibitem{Graf}
M. J. Graf, I. K. Jeong, D. L. Starr and R. H. Heffner,
Phys. Rev. B {\bf 68}, 064305 (2003).
\bibitem{Jeong2}
I.-K. Jeong, R.H. Heffner, M.J. Graf, and S. J. L. Billinge,
Phys. Rev. B {\bf 67}, 104301 (2003).
\bibitem{Booth}
C.H. Booth, F. Bridges, E.D. Bauer, G.G. Li, J.B. Boyce, T. Claeson,
C.W. Chu, and Q. Xiong, Phys. Rev. B {\bf 52}, R15 745 (1995).
\bibitem{Chung}
J.S. Chung and M.F. Thorpe, Phys. Rev. B {\bf 55}, 1545 (1997).



\end{thebibliography}
\end{document}